# Low-dimensional optical chirality in complex potentials

SUNKYU YU,[1,†] HYUN SUNG PARK,[2,†] XIANJI PIAO,[1] BUMKI MIN,[2] NAMKYOO PARK,[1,*]

[1]*Photonic Systems Laboratory, Department of Electrical and Computer Engineering, Seoul National University, Seoul 08826, Korea*
[2]*Department of Mechanical Engineering, Korea Advanced Institute of Science and Technology (KAIST), Daejeon 305-751, Korea*
*\*Corresponding author: [nkpark@snu.ac.kr](nkpark@snu.ac.kr)*
*†These authors contributed equally to this work.*

**Chirality is a universal feature in nature, as observed in fermion interactions and DNA helicity. Much attention has been given to chiral interactions of light, not only regarding its physical interpretation but also focusing on intriguing phenomena in excitation, absorption, refraction, and topological phase. Although recent progress in metamaterials has spurred artificial engineering of chirality, most approaches are founded on the same principle of the mixing of electric and magnetic responses. Here we propose nonmagnetic chiral interactions of light based on low-dimensional eigensystems. Exploiting the mixing of amplifying and decaying electric modes in a complex material, the low-dimensionality in polarization space having a chiral eigenstate is realized, in contrast to 2-dimensional eigensystems in previous approaches. The existence of optical spin black hole from low-dimensional chirality is predicted, and singular interactions between chiral waves are confirmed experimentally in parity-time-symmetric metamaterials.**

## 1. INTRODUCTION

Complex potentials that violate the Hermitian condition have been treated restrictively to describe nonequilibrium processes [1]. However, since the pioneering work of Bender [2], it has been widely accepted that the condition of parity-time (PT) symmetry allows real eigenvalues, even in complex potentials with non-Hermitian Hamiltonians. The concept of PT symmetry has opened a pathway for handling complex potentials, overcoming traditional Hermitian restrictions and stimulating the field of complex quantum mechanics [3]. In contrast to Hermitian potentials, PT-symmetric potentials support regimes in which some eigenvalues are complex [2]. Accordingly, the phase of eigenvalues can be divided into a real and a complex regime, with a border line called the exceptional point (EP) [2,4,5] marking the onset of PT symmetry breaking.

Substantial researches have focused on optical analogues of PT-symmetric dynamics [4-12]. Based on the equivalence of the Schrodinger and paraxial-wave equations, the classical simulation of complex quantum mechanics has been tested [11]. Exotic behaviors of light have also been implemented in PT-symmetric potentials: asymmetric modal conversion [8,13,14], abnormal beams [4,12], unidirectional invisibility [15], and PT-symmetric resonances [6,7,10]. These phenomena originate from complex eigenstates near the EP, in relation to their skewness [8,14] and unidirectional modal conversion [4,13]. Although there has been an effort to observe PT-symmetric dynamics in polarization space [16] as well, the previous work has focused only on the observation of PT-symmetric phase transition of eigenstates around the EP, lacking the in-depth investigation in the context of chiral 'light-matter interactions': handedness-dependent optical phenomena through the interactions with two-dimensional (2D) or three-dimensional (3D) chiral structures. For example, because the PT-symmetric metamaterial in ref. [16] was considered as a meta-'surface' between air and substrate, the studies for the evolution of propagating waves along PT-symmetric 'bulk' materials, for the increase of singular dynamics at the EP, and for the incidences with arbitrary polarizations have been neglected.

In this paper, generalizing light interactions with PT-symmetric electrical materials, we investigate a new class of 2D chiral 'bulk materials' with complex potentials, which possess the low-dimensional (1D) eigensystem and can be applied to exhibit the design of arbitrary singular polarization state on the Poincare sphere. In contrast to previous approaches [17-24] for chiral optical materials, we derive chiral interactions from the mixing of amplifying and decaying electric responses, not from the mixing of electric and magnetic responses. We demonstrate that the dimensionality of PT-symmetric chiral system is reduced to one at the EP, leading to a perfectly singular modal helix. This provides a pathway toward chiral light-matter interactions fundamentally distinct from conventional optical chirality [17-20,23,24], all of them based upon 2-dimensional eigensystems. Unique properties arising from the low-dimensionality are experimentally demonstrated in high-index metamaterials. We also show the convergence of arbitrarily-polarized incidences to a single chiral eigenstate without any reflections, realizing an optical spin black hole.

## 2. EIGENSTATES IN PT-SYMMETRICALLY POLARIZED MATERIAL

First, we assume a homogeneous and anisotropic optical material that is PT-symmetric with respect to the $(\mathbf{y} \pm \mathbf{z})/2^{1/2}$ axis. Because the permittivity tensor corresponds to the Hamiltonian for planewaves in polarization space (see Supplementary Note 1), it is necessary to define the permittivity tensor satisfying PT-symmetry condition for potential of the form $V(x) = V^*(-x)$. The permittivity tensor, satisfying the necessary condition of PT symmetry [2] ($\boldsymbol{\varepsilon}(\mathbf{r}) = \boldsymbol{\varepsilon}^*(-\mathbf{r})$), is characterized as

$$\boldsymbol{\varepsilon}_r = \begin{pmatrix} \varepsilon_{r0} & 0 & 0 \\ 0 & \varepsilon_{r0} + i\varepsilon_{i0} & \varepsilon_{\kappa 0} \\ 0 & \varepsilon_{\kappa 0}^* & \varepsilon_{r0} - i\varepsilon_{i0} \end{pmatrix}, \quad (1)$$

where $\varepsilon_{r0}$, $\varepsilon_{i0}$, and $\varepsilon_{\kappa 0}$ have real values for a nonmagnetic material (for $\varepsilon_{i0} \geq 0$, $y$: gain axis, $z$: loss axis), and $\mathbf{r}$ and $-\mathbf{r}$ satisfy the mirror (or parity) symmetry with respect to $(\mathbf{y} \pm \mathbf{z})/2^{1/2}$. For a planewave propagating along the $x$-axis ($E_y$, $E_z$), there exist two eigenstates, each with an eigenvalue (or effective permittivity) $\varepsilon_{\text{eig}1,2} = \varepsilon_{r0} \pm \lambda_{\text{PT}}$ and a corresponding eigenstate $\mathbf{v}_{\text{eig}1,2} = \eta_{1,2} \cdot (\varepsilon_{\kappa 0}, -i\varepsilon_{i0} \pm \lambda_{\text{PT}})^T$, where $\eta_{1,2} = [1/(|\varepsilon_{\kappa 0}|^2 + |-i\varepsilon_{i0} \pm \lambda_{\text{PT}}|^2)]^{1/2}$, and $\lambda_{\text{PT}} = (\varepsilon_{\kappa 0}^2 - \varepsilon_{i0}^2)^{1/2}$ is the interaction parameter [4] defining the EP ($\lambda_{\text{PT}} = 0$, Supplementary Note 1). The eigenstates $\mathbf{v}_{eig1,2}$ are nonorthogonal ($\mathbf{v}_{eig1}\cdot\mathbf{v}_{eig2}^* \neq 0$) except for Hermitian potentials ($\varepsilon_{i0} = 0$), which naturally derives intermodal coupling [8] between $\mathbf{v}_{eig1,2}$. Because the eigenstates $\mathbf{v}_{eig1,2} = \eta_{1,2}\cdot(\varepsilon_{k0}, -i\varepsilon_{i0} \pm \lambda_{PT})^T$ have elliptical polarizations in general, and the handedness of the elliptical polarizations is left handed, the PT-symmetric potential becomes naturally chiral, favoring left CP (LCP, $\mathbf{v}_L = (1/2)^{1/2}\cdot(1, -i)^T$) mode.

Figure 1a (1b) presents the real (imaginary) part of the effective permittivity $\varepsilon_{eig1,2}$ for PT-symmetric systems $d$ to $h$ of different imaginary potential $\varepsilon_{i0}$. Similar to other PT-symmetric potentials [4-12], the variation of $\varepsilon_{i0}$ derives the generic square-root curve from the definition of $\lambda_{PT}$ and results in the apparent phase transition of eigenvalues from the real to complex phase across the EP (point $f$, $\varepsilon_{i0} = \varepsilon_{k0}$ for $\lambda_{PT} = 0$). The normalized density of optical chirality [21,22] $\chi/(\beta_{r0}U_e)$ for each eigenstate in $d$-$h$ is shown in Fig. 1c (see Appendix A for the calculation of $\chi$, $U_e = |\mathbf{E}|^2$ and $\beta_{r0} = \varepsilon_{r0}^{1/2}\cdot 2\pi/\Lambda_0$), along with the corresponding profiles of eigenpolarizations (Figs 1d-1h).

For the Hermitian case (Fig. 1d), the eigenstates take linear polarizations, constituting 2-dimensional orthogonal bases. As $\varepsilon_{i0}$ increases (Fig. 1e), the eigenstates begin to converge and take the left-handed chiral form of elliptical polarizations, with nonorthogonality ($\mathbf{v}_{eig1}\cdot\mathbf{v}_{eig2}^* \neq 0$) and increased chirality (Fig. 1c, $d\rightarrow f$). At the EP (Fig. 1f), two chiral eigenstates coalesce to a LCP basis with the reduced geometric multiplicity ($\varepsilon_{eig1} = \varepsilon_{eig2}$). This low-dimensional existence of a chiral eigenstate forming a *modal helix*, which is distinguished from the structural helix of electric and magnetic mixing [18,23], yields perfect chirality with a pure handedness (Fig. 1c, point $f$). After the EP (Figs 1g,1h), the 2-dimensional eigensystem is recovered with decreased chirality (Fig. 1c, $f\rightarrow h$) as each eigenstate is saturated to a linear mode ($y$, amplifying; $z$, decaying). Note that the handedness of the chiral eigenstates can be reversed by changing the sign of $\varepsilon_{k0}$ (Fig. 1c, orange for left- and blue for right-handedness). For completeness, the imperfect PT symmetry is also investigated ($Re[\varepsilon_y] \neq Re[\varepsilon_z]$ or $Im[\varepsilon_y] \neq -Im[\varepsilon_z]$) with respect to the modal chirality (Supplementary Note 2), showing the experimental tolerance and gauge-transformed (active and passive) PT symmetry.

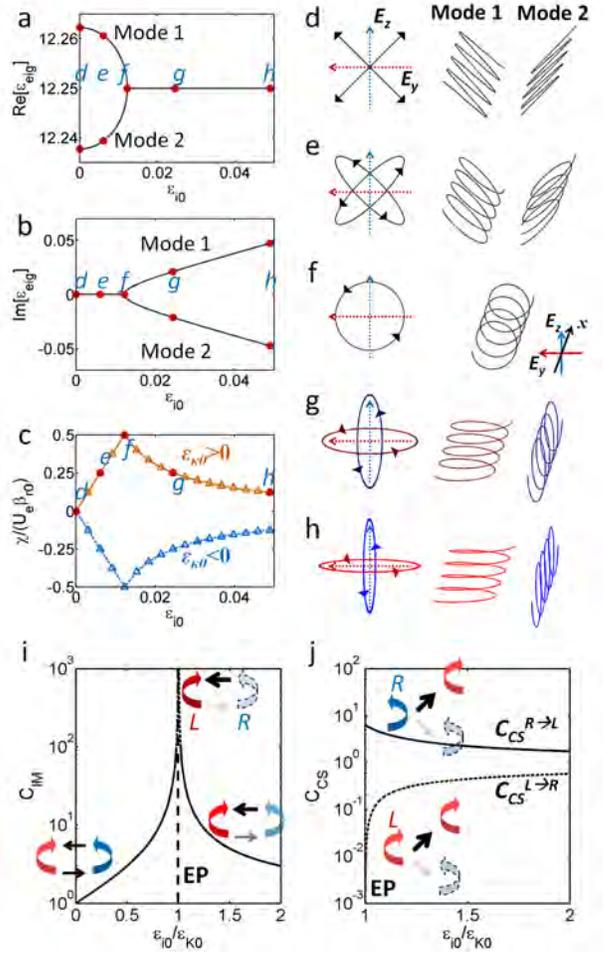

**Fig. 1. Eigenvalues and spatial evolution of eigenstates in PT-symmetric chiral material.** The real and imaginary parts of the effective permittivity $\varepsilon_{eig1,2}$ are shown in **a** and **b** with respect to $\varepsilon_{i0}$. **c.** The density of chirality $\chi$, normalized by the product of the electric field intensity $U_e$ and $\beta_{r0}$ (orange: $\varepsilon_{k0}= \varepsilon_{r0}/10^3 > 0$, blue: $\varepsilon_{k0}= -\varepsilon_{r0}/10^3 < 0$, line: eigenstate 1, symbol: eigenstate 2). **d-h**. Spatial evolution of eigenstates corresponding to points *d-h* marked in **a-c**. **d**: $\varepsilon_{i0} = 0$, **e**: $0 < \varepsilon_{i0} < \varepsilon_{k0}$, **f**: $\varepsilon_{i0} = \varepsilon_{k0}$, **g,h**: $\varepsilon_{i0} > \varepsilon_{k0}$. The red and blue arrows represent the axes of $E_y$ (amplifying mode) and $E_z$ (decaying mode). At the EP *f*, the complex eigenstate has the singular form of a modal helix. **i**. $C_{IM}$ and **j**. $C_{CS}$ as functions of ($\varepsilon_{i0}/\varepsilon_{K0}$). $\varepsilon_{r0} = 12.25$ for **a-h**, and $\varepsilon_{r0}=6.5$ for **j**. $\varepsilon_{k0} = \varepsilon_{r0}/10^3 > 0$ for **a, b, d-h,** and **j**. $L_{eff}=10^3$ for **b**.

## 3. CHIRAL INTERACTIONS IN PT-SYMMETRICALLY POLARIZED MATERIAL

To examine the role of chiral eigenstates in PT-symmetric material, here we focus on the 'chiral interaction' with singularity, extending the discussion in ref. [16] which focused on the eigenstate itself. Firstly, we study the modal transfer between CP modes through propagation. Utilizing the eigenstates and eigenvalues from Eq. (1) and employing the CP bases of $\mathbf{v}_{RL} = (1/2)^{1/2}\cdot(1, \pm i)^T$, the transfer relation between the incident $E_{inc} = (E_{RI}, E_{LI})_{CP}^T$ and transmitted field $E_{trn} = (E_{RT}, E_{LT})_{CP}^T$ is written as $\mathbf{E}_{trn} = \mathbf{M}_{PT}\mathbf{E}_{inc}$ using only structural (propagation distance $d$) and material ($\varepsilon_{r0}$, $\varepsilon_{i0}$, and $\varepsilon_{k0}$) parameters (Supplementary Note 3).

A closer investigation of the transfer matrix $\mathbf{M}_{PT}$ in Supplementary Note 3 provides a straightforward understanding of chiral interactions in PT-symmetric potentials. First, the inequality between off-diagonal terms ($|t_{R\rightarrow L}| > |t_{L\rightarrow R}|$) leads to an asymmetric modal conversion between the right- and left-CP (RCP and LCP) modes. Because the self-evolutions of CP modes are identical ($t_{R\rightarrow R} = t_{L\rightarrow L}$), the chiral response of the system is governed by the *intermodal chirality* $C_{IM}$ as

$$C_{\text{IM}} = \left| \frac{t_{\text{R}\to\text{L}}}{t_{\text{L}\to\text{R}}} \right| = \left| \frac{\varepsilon_{\kappa 0} + \varepsilon_{\text{i}0}}{\varepsilon_{\kappa 0} - \varepsilon_{\text{i}0}} \right| = \left| \frac{1 + \frac{\varepsilon_{\text{i}0}}{\varepsilon_{\kappa 0}}}{1 - \frac{\varepsilon_{\text{i}0}}{\varepsilon_{\kappa 0}}} \right|. \quad (2)$$

which is obtained in Supplementary Note 3.

Note that the intermodal chirality $C_{\text{IM}}$ is solely determined by the ratio of $\varepsilon_{\text{i}0}$ and $\varepsilon_{\kappa 0}$, directly related to the degree of PT symmetry, i.e., $\lambda_{\text{PT}} = (\varepsilon_{\kappa 0}^2 - \varepsilon_{\text{i}0}^2)^{1/2}$. Accordingly, at the EP ($\varepsilon_{\text{i}0} = \varepsilon_{\kappa 0}$), a one-way chiral conversion $C_{\text{IM}} \to \infty$ from the RCP to LCP mode is achieved (Fig. 1i), as expected from the reduction of the eigensystem to a 1-dimensional LCP eigenstate (Fig. 1f). Before and after the EP, $C_{\text{IM}}$ decreases (Fig. 1i) due to two elliptically-polarized eigenstates (Figs 1e,1g). Note that this low-dimensional PT-chiral system ($t_{\text{R}\to\text{R}} = t_{\text{L}\to\text{L}}$ and $t_{\text{R}\to\text{L}} \gg t_{\text{L}\to\text{R}} \sim 0$) is fundamentally distinct from conventional chirality based on a 2-dimensional Hermitian eigensystem of circular birefringence ($t_{\text{R}\to\text{R}} \neq t_{\text{L}\to\text{L}}$) and zero intermodal coupling ($t_{\text{L}\to\text{R}} = t_{\text{R}\to\text{L}} = 0$) [18,20,22,23,25].

When the intermodal chirality $C_{\text{IM}}$ defines the unique origin of low-dimensional chirality in PT-symmetric material, the *strength* of the chiral conversion $C_{\text{CS}}$ (Fig. 1j) is determined by the competition between the intermodal transfer and the self-evolution ($C_{\text{CS}}^{\text{R}\to\text{L}} = |t_{\text{R}\to\text{L}} / t_{\text{R}\to\text{R}}|$, $C_{\text{CS}}^{\text{L}\to\text{R}} = |t_{\text{L}\to\text{R}} / t_{\text{L}\to\text{L}}|$, see Supplementary Note 4 for details). In agreement with the observations made for $C_{\text{IM}}$, $C_{\text{CS}}^{\text{R}\to\text{L}}$ is always larger than $C_{\text{CS}}^{\text{L}\to\text{R}}$, resulting in LCP-favored chiral conversion (Fig. 1j). Near the EP (see Supplementary Note 4, $C_{\text{CS}}^{\text{R}\to\text{L}} \sim 2\pi L_{\text{eff}}\cdot(\varepsilon_{\text{i}0}/\varepsilon_{\text{r}0})$ and $C_{\text{CS}}^{\text{L}\to\text{R}} \sim 0$), the chiral conversion becomes unidirectional, and its strength is solely determined by the material and structural parameters: $\varepsilon_{\text{i}0}/\varepsilon_{\text{r}0}$ and effective interaction length $L_{\text{eff}} = n_{\text{eff}}d/\Lambda_0$ where $n_{\text{eff}}$ is the effective index of the structure. We note that while the magnitude of the gain and loss $\varepsilon_{\text{i}0}$ contributes both to intermodal chirality $C_{\text{IM}}$ and chiral conversion $C_{\text{CS}}$, the achievement of large $n_{\text{eff}}$, e.g., with unnaturally high-index metamaterials [26], enables strong chiral conversion (or large $C_{\text{CS}}$) in the 'bulk', overcoming the material restriction on $\pm\varepsilon_{\text{i}0}$.

Figures 2a and 2b show the chirality of the transmitted wave for RCP and LCP incidences. While $\varepsilon_{\text{i}0}/\varepsilon_{\text{r}0}$ determines the regime of chiral transfers (oscillatory for $\varepsilon_{\text{i}0}/\varepsilon_{\text{r}0} < 1$, and LCP-favored for $\varepsilon_{\text{i}0}/\varepsilon_{\text{r}0} \geq 1$), $L_{\text{eff}}$ describes the effect of the interaction length of PT-symmetric chiral materials. For large $L_{\text{eff}}$, strong LCP chirality from the one-way chiral conversion is apparent near the EP, which emphasizes the role of the singularity. This chiral singularity forms the *optical spin black hole* (the south pole of the Poincaré sphere, Fig. 2c), to where all the states of polarization (SOP) converge (Supplementary Note 5 for details, and Supplementary Movie 1 for the spin black hole behavior at the EP), differentiating the analysis based on low-dimensional chirality from conventional chirality [17-24] or the analysis of the 'eigenstate' singularity in PT-symmetric metasurfaces [16]. Because the eigenstate of the polarization singularity can be designed systematically (e.g. see Supplementary Note 6 for a low-dimensional eigenstate of linear polarization (LP)), the optical spin black hole can be achieved in the entire polarization space of the Poincaré sphere, by combining the imaginary potential for linear and circular bases. We also reveal another unique property of low-dimensional chirality in Supplementary Note 7, chirality *reversal* for LP incidences (Supplementary Movie 2 for LP incidence), which is impossible for the optical activity in conventional chiral materials [27].

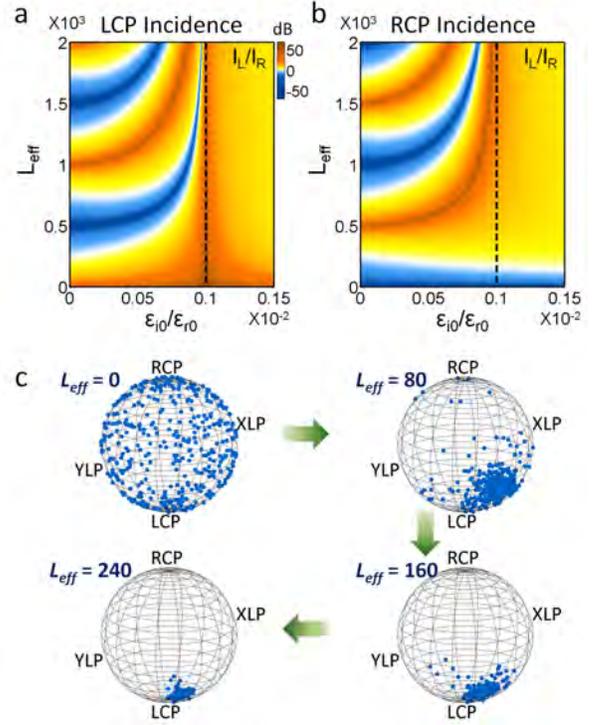

**Fig. 2. Chiral dynamics within PT-symmetric optical material.** The output power ratio of LCP over RCP ($I_L/I_R = |E_{\text{LT}}/E_{\text{RT}}|^2$ in dB) is shown for the case of **a**. LCP and **b**. RCP incidence as a function of the imaginary permittivity ($\varepsilon_{\text{i}0}/\varepsilon_{\text{r}0}$) and the interaction length ($L_{\text{eff}} = \varepsilon_{\text{r}0}^{1/2}\cdot d/\Lambda_0$). The black dotted lines in **a** and **b** represent the EPs, where the dimensionality reduces to one. **c**. LCP-convergent spin black hole dynamics on the Poincaré sphere at the EP, demonstrated with randomly polarized incidences. The interaction lengths are $L_{\text{eff}} = 0$, 80, 160, and 240, clockwise from the upper left. The movie is shown in Supplementary Movie 1. All the results are based on the transfer matrix method. $\varepsilon_{\text{r}0}=6.5$, and $\varepsilon_{\kappa 0}=\varepsilon_{\text{r}0}/10^3$ in **a** and **b,** $\varepsilon_{\kappa 0}=\varepsilon_{\text{r}0}/200$ in **c**.

## 4. GIANT CHIRAL CONVERSION IN THE RESONANT STRUCTURE

To achieve a large chiral conversion within a compact footprint, a resonant structure for the effective increase of interaction length $L_{\text{eff}} = \varepsilon_{\text{r}0}^{1/2}\cdot d/\Lambda_0$ can be considered (Fig. 3a). The resonator is composed of PT-symmetric anisotropic material at the EP (length $d = 834$ nm, same parameters as those of Fig. 2), sandwiched between two metallic mirrors (of thickness $\delta$). Here, S-matrix analysis [28] is utilized to calculate the frequency-dependent transmission, reflection, and field distributions inside the resonator. In detail, while the fields of the background air and mirrors are expressed as the linear combinations of $y$- and $z$-orthogonal bases, the field inside the PT-symmetric material is expressed by the nonorthogonal bases of $\mathbf{v}_{\text{eig}1,2}$, as $\mathbf{E}(x) = E_{\text{eig}1}^+\cdot\mathbf{v}_{\text{eig}1}\cdot exp(-i\beta_1\cdot x) + E_{\text{eig}1}^-\cdot\mathbf{v}_{\text{eig}1}\cdot exp(i\beta_1\cdot x) + E_{\text{eig}2}^+\cdot\mathbf{v}_{\text{eig}2}\cdot exp(-i\beta_2\cdot x) + E_{\text{eig}2}^-\cdot\mathbf{v}_{\text{eig}2}\cdot exp(i\beta_2\cdot x)$, including forward (+) and backward (−) components. From the continuity condition of the electric field across the boundary ($\mathbf{E}$ and $\partial_x\mathbf{E}$), we derive the S-matrix relation of $(E_{0y}^+, E_{0z}^+, E_{\text{I}y}^-, E_{\text{I}z}^-)^T = \mathbf{S}\cdot(E_{\text{I}y}^+, E_{\text{I}z}^+, E_{0y}^-, E_{0z}^-)^T$. For mirrors with thicknesses of $\delta = 40$, 50, 60, or 70 nm, the obtained $Q$ values of the resonators are 620, 1500, 3600, and 8200, respectively.

Figures 3b and 3c show the S-matrix calculated power ratio of LCP over RCP of a transmitted and reflected wave for the forward $y$-linear polarized incidence ($E_{\text{I}y}^+ = 1$). Also shown in Fig. 3d are the S-matrix-calculated wave evolutions at the on-resonance condition of the 3/2 wavelength. Enhanced by the chiral standing wave in the resonator, a giant LCP-favored chirality of the transmitted wave is observed (Fig. 3b, $I_L/I_R = 20$ dB within $L_{\text{eff}} = 1.4$ at $Q = 8200$; to compare, for the non-resonant structure, [$I_L/I_R$ = 0.08 dB, $L_{eff}$ = 1.4] and [$I_L/I_R$ = 20 dB, $L_{eff}$ = 1450]). It is also notable that, in contrast to the non-resonant structure where the reflection is absent, *pure chiral reflection* (RCP-only) results from the backward-propagating RCP waves inside the resonator (Fig. 3c).

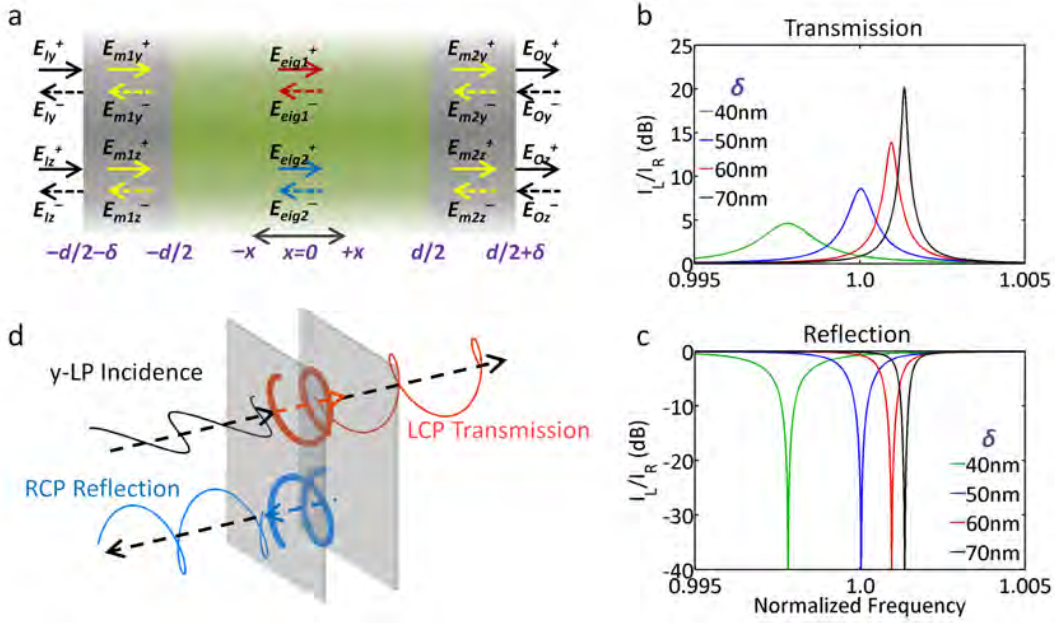

**Fig. 3. Giant chiral conversion through the resonant structure. a.** Schematics of the chiral resonator for the S-matrix analysis (green: PT-symmetric anisotropic material of $d$ = 834 nm, $\varepsilon_{r0}$ = 6.5 and $\varepsilon_{k0} = \varepsilon_{i0} = \varepsilon_{r0}/1000$; grey: metallic mirrors, $\varepsilon_{metal}$ = −100, $\Lambda_0$ = 1500 nm). $L_{eff}$ = 1.4. The power ratios of LCP over RCP in the **b.** transmitted and **c.** reflected wave for different mirror thicknesses. **d.** S-matrix-based spatial evolutions of waves through the resonator. Arrows denote the propagating direction of each wave.

## 5. LOW-DIMENSIONAL CHIRAL METAMATERIAL

We now investigate the experimental realization of the PT-symmetric chiral medium, focusing on the observation of the one-way chiral conversion $C_{IM}$ as a clear and direct evidence of the low-dimensionality. To achieve the complex anisotropic permittivity of Eq. (1) with isotropic media, we employ the platform of metamaterials; which enables the use of local material parameter from the subwavelength structure and the implementation of gauge-transformed [29] PT-symmetry in a passive manner (Supplementary Note 2) through the designer permittivity.

Figure 4 shows the realization of low-dimensional chiral metamaterial, achieved by transplanting the ideal point-wise anisotropic permittivity (Fig. 4a) into the subwavelength structure (see Appendix B and C for the fabrication and THz measurement). We emphasize that the chiral conversion $C_{CS}^{R \to L} = |t_{R \to L} / t_{R \to R}| \sim 2\pi L_{eff}(\varepsilon_{i0}/\varepsilon_{r0})$ is directly proportional to $n_{eff} = (\mu_r \varepsilon_r)^{1/2}$. Considering that most of capacitive metamaterials (including split-ring resonators in [16]) have strong diamagnetic behavior ($\mu_r \ll 1$) [30] hindering the realization high effective index, we employ the I-shaped metamaterial which provides an ultrahigh permittivity [26] ($\varepsilon_r \gg 1$) and suppressed magnetic moments ($\mu_r \sim 1$) [30]; achieving *strong* and purely *electrical* light-matter interaction in the THz regime. We also note that the multilayer extension can be obtained for I-shaped structures [26], and the operation condition of high effective index and purely electrical response can be satisfied in the optical regime as well, for example by utilizing hyperbolic metamaterials [31].

Because the I-shaped patch supports effective permittivity following the Lorentz model [26], its spectral response (Fig. 4b) is divided into dielectric ($Re[\varepsilon] \geq 0$) and metallic ($Re[\varepsilon] < 0$) states, both of which can be subdivided into low-loss ($|Re[\varepsilon]| \gg |Im[\varepsilon]|$) and high-loss ($|Re[\varepsilon]| \sim |Im[\varepsilon]|$) regimes. To compose the PT-symmetric *polar* metamaterial, we utilize the low- and high-loss regimes for $y$ and $z$ polarization, respectively.

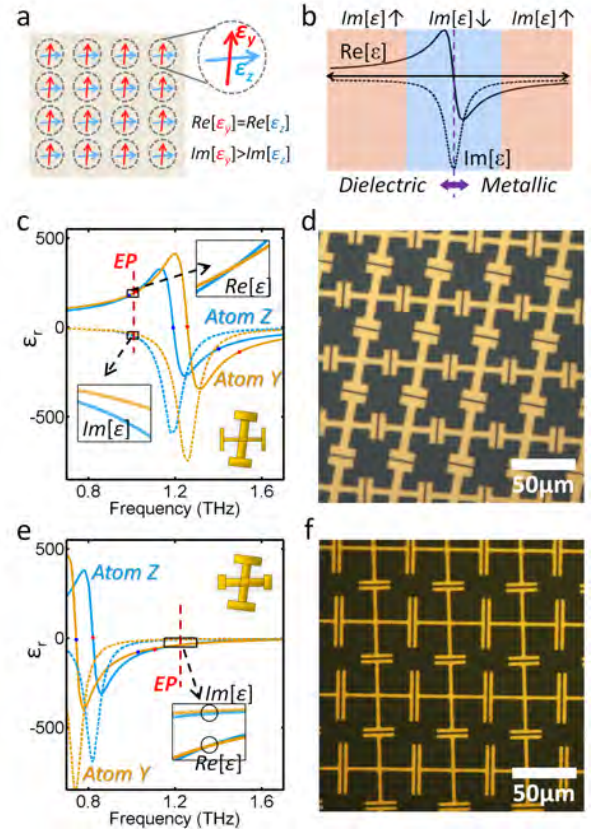

**Fig. 4. Chiral polar metamaterial for low-dimensional chirality. a.** Schematics of a PT-symmetric, point-wise anisotropic permittivity material for Eq. (1). **b.** Lorentz model for an I-shaped patch with different material regimes. The effective anisotropic permittivity of chiral metamaterials (Supplementary Note 8): **c.** the dielectric realization with propagating mode, and **e.** the metallic realization with evanescent mode. The fabricated samples of a chiral metamaterial are shown in **d.** and **f.**, respectively. Insets of **c** and **e** are the expanded images of the real and imaginary parts near the EP (red dotted line). The width of each polarized patch in **c,d** is set unequally to $w_y$ = 5.5 μm and $w_z$ = 7.5 μm, and the other structural parameters are $g$ = 1.0 μm, $L$ = 34.5 μm, $a$ = 20.5 μm, $t$ = 100 nm, and $d$ = 2 μm. The arm length of each polarized patch in **e,f** is set unequally to $a_y$ = 25 μm and $a_z$ = 40 μm, and the other structural parameters are $g$ = 1.5 μm, $L$ = 50 μm, $w$ = 3.0 μm, $t$ = 100 nm, and $d$ = 2 μm. See Supplementary Fig. 5a for the definition of structural parameters.

Figures 4c,4d (Figs 4e,4f) show the experimental realization of PT-symmetric *polar* metamaterial, using a dielectric (metallic) state (see Supplementary Note 8). Figures 4c and 4e show anisotropic permittivity $\varepsilon_y$ and $\varepsilon_z$ (at $\theta$ = 0 in Supplementary Fig. 5a) for dielectric and metallic realizations (Figs 4d and 4f, respectively). These figures show the spectral overlap between low- and high-loss regimes for the $y$- and $z$-axis patches and the existence of spectral EP (with $Re[\varepsilon_y]$ = $Re[\varepsilon_z]$ and $Im[\varepsilon_y] \neq Im[\varepsilon_z]$, Supplementary Note 2). Note that our single-layered structure embedded in a subwavelength-thick polyimide ($t < \lambda_0/100$) can be considered as a homogenized metamaterial [26]. To introduce the coupling $\varepsilon_{\kappa 0}$, we apply the oblique alignment with a tilted angle $\theta$ (Supplementary Fig. 5a).

By changing $\theta$ (the coupling $\varepsilon_{\kappa 0}$) in the fabricated sample, now we measure the $\theta_{EP}$, where the one-way chiral conversion of singularity occurs with $\varepsilon_{\kappa 0}$ = $|Im[\varepsilon_y-\varepsilon_z]|$. The experimentally measured intermodal chirality $C_{IM}(\theta,\omega)$ is shown each for dielectric (Fig. 5a) and metallic (Fig. 5b) state realization, in good agreement with the COMSOL simulation (Figs 5c and 5d, respectively). In both samples, $\theta_{EP}$ of singularity in ($\theta,\omega$) space, satisfying the one-way chiral conversion for the sensitive EP are observed (cross point of black dotted lines). For the dielectric state metamaterial with propagating waves inside, the large $C_{IM}$ = 17.4 dB is observed at $\theta_{EP}$ = 2.0°, and $C_{IM}$ = 16.4 dB is observed at $\theta_{EP}$ = 1.6° in the design in the metallic state with evanescent waves inside the metamaterial. Despite the different propagating features of the two regimes, the EP design derives the chiral interaction of light in terms of 'one-way chiral conversion' for both regimes, confirming the role of low-dimensionality with a singular chiral eigenstate. Note that the separated $y$ and $z$ local modes that are highly-concentrated within the gaps are well-converted to a single planewave-like beam due to the deep-subwavelength scale of the structure, conserving the pure spin angular momentum of light without additional orbital angular momentum.

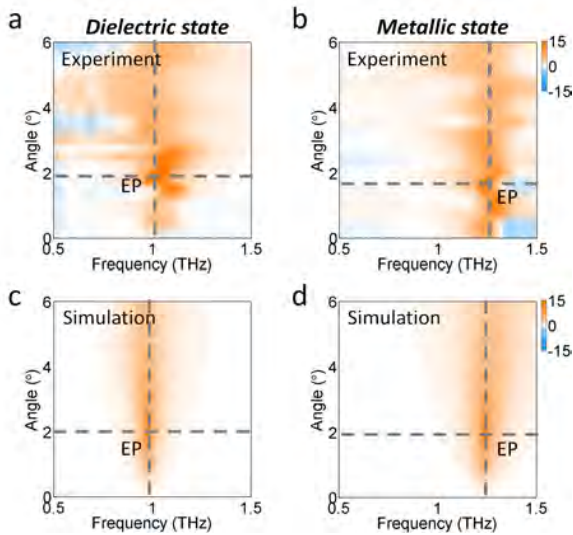

**Fig. 5. Observation of EP in chiral polar metamaterials.** The experimental (**a,b**) and the simulated (**c,d**) results of $C_{IM}$ are shown in a spectral regime for dielectric (**a,c**) and metallic (**b,d**) realizations. Dotted lines represent the condition of EPs in spectral and $\theta$ domains. All simulated results were obtained using COMSOL Multiphysics.

## 6. LOW-DIMENSIONAL MODAL HELIX IN GUIDED-WAVE STRUCTURES

While ref. [16] has been focused on the singular 'scattering' from PT-symmetric meta-'surfaces' which can be analyzed through the scattering matrix, the impact of 'eigenstates' defined the system Hamiltonian is critical for propagating waves along the 'bulk'. Extending the discussion to the guided-wave and the optical frequency, we propose a modal helix in an optical waveguide platform utilizing isotropic materials. The point-wise permittivity (Fig. 4a) is transplanted into the complex-strip waveguide as a passive form (Fig. 6a), where the lossy Ti layer (grey region, thickness $t_{Ti}$) under a lossless Si-strip waveguide imposes the selective decay of the $z$ mode, which is well-separated from the $y$ mode (Fig. 6b, $\varepsilon_y$ for low-loss and $\varepsilon_z$ for high-loss). The structural parameters are designed to satisfy $Re[\varepsilon_y]$ = $Re[\varepsilon_z]$, and the coupling $\varepsilon_{yz}$ is achieved with the deviation $\Delta$, which breaks the orthogonality between the $y$ and $z$ polarized modes. It is worth mentioning that the sign change of $\varepsilon_{\kappa 0}$ (=$\varepsilon_{yz}$) can also be controlled by the mirror offset of $-\Delta$ for the deterministic control of the handedness (Fig. 1c). Therefore, the chirality of the proposed modal helix has *directionality* from the sign reversal of $\Delta$ for the backward (**-x**) view, which is absent in the structural helicity [18,23].

Figures 6c and 6d show the COMSOL-calculated modal chirality and the difference between eigenvalues as a function of the structural parameters ($t_{Ti}$ and $\Delta$). Because the control of the Ti layer alters the complex part of $\varepsilon$, the two structural parameters provide three degrees of freedom ($Re[\varepsilon]$, $Im[\varepsilon]$, and $\varepsilon_{\kappa 0}$), resulting in the single EP in the 2D parameter space ($t_{Ti}$ = 19 nm, $\Delta$ = 91 nm, $I_L/I_R$ = 21 dB, and 18 dB for modes 1 and 2). The finite modal chirality (~20 dB) originates in the separated intensity profiles of the $y$ and $z$ modes (Fig. 6b), resulting in the non-uniform local chirality (82 dB maximum, Fig. 6e). Therefore, the chiral guided-wave includes the additional orbital angular momentum from the varying wave-front, in contrast to the case of the subwavelength structure (Fig. 4), which supports a planewave. With an experimentally accessible geometry ($I_L/I_R \geq$ 10 dB in $\Delta$ = 80~100 nm) and the coalescence of eigenmodes (Fig. 6d), the complex-strip waveguide will be an ideal building block for chiral guided-wave devices and for the utilization of active materials such as GaInAsP. It is also worth mentioning that our approach based on the guided-wave platform can derive the low-dimensional chirality through the simple spatial displacement of the lossy dielectric waveguide, without the use of molecular designs for low-loss and high-loss components [16].

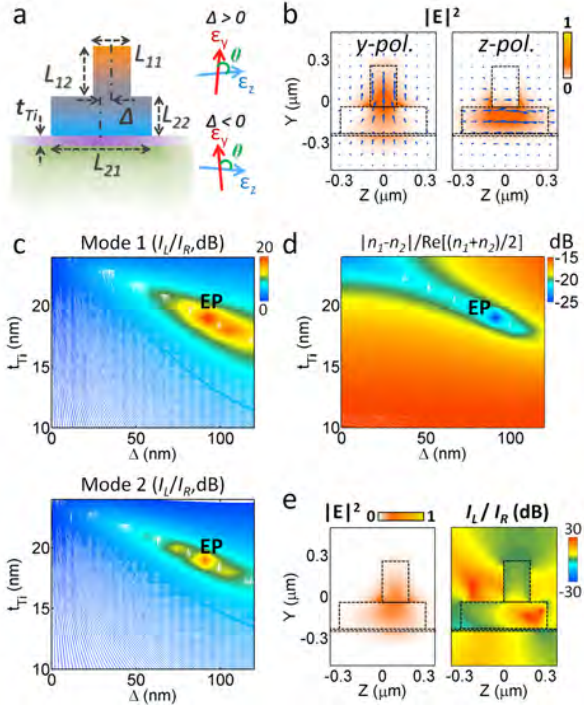

**Fig. 6. Chiral waveguides supporting bases of modal helices. a**. Cross sections of a complex-strip waveguide using isotropic materials (graded color: silicon; purple: titanium; green: silica substrate; graded color represents the effective loss by the titanium layer). The lossless silicon ($\varepsilon_{Si}$=12.1) is assumed to compose the strip structure on top of the lossy titanium layer ($\varepsilon_{Ti}$=1.66-$i$·30.1) above a silica substrate ($\varepsilon_{SiO2}$=2.07), supporting both a low-loss $y$-polarized mode and a high-loss $z$-polarized mode. The complex-strip waveguide satisfies PT symmetry based on the gauge transformation ($Re[\varepsilon_y]=Re[\varepsilon_z]$ and $Im[\varepsilon_z]<Im[\varepsilon_y]<0$). The effect of the loss can be controlled by changing the depth of the titanium layer. The red and blue arrows describe a corresponding point-wise anisotropic permittivity. **b**. The intensity profile and the polarization (in arrows) of the eigenmodes for the structure ($\varepsilon_{yz}$=0, $\Delta$=0). **c** shows the modal chirality by $I_L/I_R$ as a function of $\Delta$ and $t_{Ti}$. **d**. The absolute value of the difference between eigenvalues as a function of $\Delta$ and $t_{Ti}$. The intensity profile and the local chirality ($I_L(y,z) / I_R(y,z)$) at the EP are shown in **e**. All results were obtained using COMSOL Multiphysics with an optical wavelength of $\Lambda_0$=1500 nm. $L_{11}$=190 nm, $L_{12}$=300 nm, $L_{21}$=620 nm, and $L_{22}$=190 nm.

## 7. CONCLUSION

We proposed and investigated a new class of optical chiral interactions based on complex potentials, without any bi-anisotropic mixing of electric and magnetic dipoles [27]. Based on the mixing of amplifying and decaying responses and on the presence of the low-dimensional eigenstate at the EP, exotic chiral behaviors of one-way CP convergence are observed, which cannot be observed in conventional chiral or gyrotropic materials. The reduced dimensionality also enables reflectionless CP generation following the dynamics of optical spin black holes, which is impossible in conventional approaches based on Hermitian elements. We also emphasize that our result, supporting the saturation of linear polarizations to a single circular polarization (Supplementary Note 7), is distinct from conventional chirality, which maintains optical activity for linear polarizations. We demonstrated the physics of low-dimensional chirality by using ultrahigh index polar metamaterial [26] with two design strategies: utilizing propagating and evanescent waves. The results prove the existence of EP and the one-way chiral conversion in the spectral regime, and the manipulation of the angular property is achieved. As an application, a chiral waveguide using isotropic materials is also proposed, which can be achieved by transplanting the point-wise anisotropic permittivity of an ideal complex-potential. Compared to previous results for polarization tunability based on detuned dipoles [32], temporal retardation [33], and singular scattering [16], our study focuses on designing target eigenstates in a chiral form, also revealing the exotic phenomenon of optical spin black hole.

Our new findings of low-dimensional chirality will pave a route toward active chiral devices [25], such as on-chip guided-wave devices for chiral lasers, amplifiers, absorbers, and switches [34] as well as complex chiral metamaterials [9] and topological phases. From the evident correlation between complex potentials and optical chirality, we can also imagine the eigenstate with nontrivial optical spins in 'non-PT-symmetric' complex potentials, by utilizing supersymmetry technique [35,36] or the inverse design of an eigenstate in disordered media [37]. Based on the general framework of non-Hermitian physics, we note that our work can be further extended using different polarization bases (Supplementary Note 6) to enable SOP collection for the arbitrary designer polarization.

## 8. APPENDIX

### A. Density of optical chirality for complex eigenstates

For the time-harmonic field of $\mathbf{E} = \mathbf{E}_0 \cdot e^{i\omega t}$ and $\mathbf{B} = \mathbf{B}_0 \cdot e^{i\omega t}$, the time-varying representation of the optical chirality density [21,22] $\chi = [\varepsilon_0\varepsilon_r \cdot \mathbf{E}(t) \cdot (\nabla \times \mathbf{E}(t)) + (1/\mu_0) \cdot \mathbf{B}(t) \cdot (\nabla \times \mathbf{B}(t))] / 2$ is simplified to the time-averaged form of $\chi = \omega \cdot Im[\mathbf{E}_0^* \cdot \mathbf{B}_0] / 2$. Because $\mathbf{E}_0 = \mathbf{v}_{eig1,2} \cdot exp(-i\beta_{1,2}x)$, where $\beta_{1,2}=2\pi\varepsilon_{eig1,2}^{1/2}/\Lambda_0$, the magnetic field $\mathbf{B}_0$ is

$$B_0 = -\frac{\eta_{1,2}\beta_{1,2}}{\omega} \cdot \begin{bmatrix} -i\varepsilon_{i0} \pm \lambda_{PT} \\ -\varepsilon_{\kappa 0} \end{bmatrix} \cdot e^{-i\beta_{1,2}x}, \quad (3)$$

From the definition of $\chi$ and the condition of weak coupling in the PT-symmetric system ($\varepsilon_{i0} \sim \varepsilon_{\kappa 0} << \varepsilon_{r0}$), the chirality density of each eigenstate before and after the EP is now expressed as

$$\chi_{1,2} = \frac{\beta_{1,2}}{2} \cdot \frac{\varepsilon_{i0}}{\varepsilon_{\kappa 0}}, \quad (4)$$

$$\chi_{1,2} = \frac{Re[\beta_{1,2}]}{2} \cdot \frac{2\varepsilon_{\kappa 0} \cdot (\varepsilon_{i0} \mp \sqrt{\varepsilon_{i0}^2 - \varepsilon_{\kappa 0}^2})}{\varepsilon_{\kappa 0}^2 + (\varepsilon_{i0} \mp \sqrt{\varepsilon_{i0}^2 - \varepsilon_{\kappa 0}^2})^2} \cdot e^{2Im(\beta_{1,2})x}, \quad (5)$$

where $exp(2 \cdot Im[\beta_{1,2}]x) = |\mathbf{E}_0|^2 \equiv U_e$ represents the amplifying and decaying electric field intensities after the EP (Fig. 1b, $U_e$ = 1 before the EP). Herein, we adopt $\chi_{1,2}/U_e$ to express the energy-normalized chirality of the eigenstates. Note that $\chi_1 \sim \chi_2$ from the condition of weak coupling (line and symbol of Fig. 1c).

### B. Fabrication process of THz chiral polar metamaterials

Serving as a flexible and vertically symmetric environment of a metamaterial, a polyimide solution (PI-2610, HD MicroSystems) was spin-coated (1 μm) onto a bare Si substrate and converted into a fully aromatic and insoluble polyimide (baked at 180°C for 30 min and cured at 350°C). A negative photoresist (AZnLOF2035, AZ Electronic Materials) was spin-coated and patterned using photolithography. Then, Au (100 nm) was evaporated on the Cr (10 nm) adhesion layer and patterned as crossed 'I'-shaped array structures via the lift-off process. Repeating the polyimide coating and curing (1 μm), single-layered metamaterials were fabricated by peeling off the metamaterial layers from the substrate.

### C. THz-TDS system for the measurement of intermodal chirality $C_{IM}$

To generate a broadband THz source, a Ti:sapphire femtosecond oscillator was used (Mai-Tai, Spectra-physics, 80 MHz repetition rate, 100 fs pulse width, 800 nm central wavelength, and 1 W average power). The pulsed laser beam was focused onto a GaAs terahertz emitter (Tera-SED, Gigaoptics). The emitted THz wave was then focused onto the samples using a 2 mm spot diameter. The propagating THz radiation was detected through electro-optical sampling using a nonlinear ZnTe crystal. The THz-TDS system has a usable bandwidth of 0.1-2.6 THz and a signal-to-noise ratio greater than 10,000:1.

**Funding**. This work was supported by the National Research Foundation of Korea (NRF) through the Global Frontier Program (GFP) NRF-2014M3A6B3063708 and the Global Research Laboratory (GRL) Program K20815000003, and the Brain Korea 21 Plus Project in 2015, which are all funded by the Ministry of Science, ICT & Future Planning of the Korean government. S. Yu was also supported by the Basic Science Research Program (2016R1A6A3A04009723) through the NRF, funded by the Ministry of Education of the Korean government.

**Acknowledgment**. We thank J. Hong for reading the manuscript and providing useful feedback.

## REFERENCES

1. N. Hatano, and D. R. Nelson, "Localization transitions in non-Hermitian quantum mechanics," Phys. Rev. Lett. **77**, 570 (1996).
2. C. M. Bender, and S. Boettcher, "Real spectra in non-Hermitian Hamiltonians having PT symmetry," Phys. Rev. Lett. **80**, 5243 (1998).
3. C. M. Bender, D. C. Brody, and H. F. Jones, "Complex extension of quantum mechanics," Phys. Rev. Lett. **89**, 270401 (2002).
4. S. Yu, D. R. Mason, X. Piao, and N. Park, "Phase-dependent reversible nonreciprocity in complex metamolecules," Phys. Rev. B **87**, 125143 (2013).
5. L. Feng, X. Zhu, S. Yang, H. Zhu, P. Zhang, X. Yin, Y. Wang, and X. Zhang, "Demonstration of a large-scale optical exceptional point structure," Opt. Express **22**, 1760-1767 (2014).
6. B. Peng, Ş. K. Özdemir, F. Lei, F. Monifi, M. Gianfreda, G. L. Long, S. Fan, F. Nori, C. M. Bender, and L. Yang, "Parity-time-symmetric whispering-gallery microcavities," Nature Phys. **10**, 394-398 (2014).
7. H. Hodaei, M.-A. Miri, M. Heinrich, D. N. Christodoulides, and M. Khajavikhan, "Parity-time–symmetric microring lasers," Science **346**, 975-978 (2014).
8. C. E. Rüter, K. G. Makris, R. El-Ganainy, D. N. Christodoulides, M. Segev, and D. Kip, "Observation of parity–time symmetry in optics," Nature Phys. **6**, 192-195 (2010).
9. R. Fleury, D. L. Sounas, and A. Alù, "Negative refraction and planar focusing based on parity-time symmetric metasurfaces," Phys. Rev. Lett. **113**, 023903 (2014).
10. B. Peng, Ş. Özdemir, S. Rotter, H. Yilmaz, M. Liertzer, F. Monifi, C. Bender, F. Nori, and L. Yang, "Loss-induced suppression and revival of lasing," Science **346**, 328-332 (2014).
11. S. Deffner, and A. Saxena, "Jarzynski Equality in PT-Symmetric Quantum Mechanics," Phys. Rev. Lett. **114**, 150601 (2015).
12. K. Makris, R. El-Ganainy, D. Christodoulides, and Z. H. Musslimani, "Beam dynamics in P T symmetric optical lattices," Phys. Rev. Lett. **100**, 103904 (2008).
13. S. Yu, X. Piao, K. Yoo, J. Shin, and N. Park, "One-way optical modal transition based on causality in momentum space," Opt. Express **23**, 24997-25008 (2015).
14. S. Yu, X. Piao, D. R. Mason, S. In, and N. Park, "Spatiospectral separation of exceptional points in PT-symmetric optical potentials," Phys. Rev. A **86**, 031802 (2012).
15. X. Zhu, L. Feng, P. Zhang, X. Yin, and X. Zhang, "One-way invisible cloak using parity-time symmetric transformation optics," Opt. Lett. **38**, 2821-2824 (2013).
16. M. Lawrence, N. Xu, X. Zhang, L. Cong, J. Han, W. Zhang, and S. Zhang, "Manifestation of PT Symmetry Breaking in Polarization Space with Terahertz Metasurfaces," Phys. Rev. Lett. **113**, 093901 (2014).
17. V. Fedotov, P. Mladyonov, S. Prosvirnin, A. Rogacheva, Y. Chen, and N. Zheludev, "Asymmetric propagation of electromagnetic waves through a planar chiral structure," Phys. Rev. Lett. **97**, 167401 (2006).
18. J. K. Gansel, M. Thiel, M. S. Rill, M. Decker, K. Bade, V. Saile, G. von Freymann, S. Linden, and M. Wegener, "Gold helix photonic metamaterial as broadband circular polarizer," Science **325**, 1513-1515 (2009).
19. Z. Li, M. Gokkavas, and E. Ozbay, "Manipulation of asymmetric transmission in planar chiral nanostructures by anisotropic loss," Advanced Optical Materials **1**, 482-488 (2013).
20. J. Pendry, "A chiral route to negative refraction," Science **306**, 1353-1355 (2004).
21. Y. Tang, and A. E. Cohen, "Optical chirality and its interaction with matter," Phys. Rev. Lett. **104**, 163901 (2010).
22. Y. Tang, and A. E. Cohen, "Enhanced enantioselectivity in excitation of chiral molecules by superchiral light," Science **332**, 333-336 (2011).
23. M. Thiel, M. S. Rill, G. von Freymann, and M. Wegener, "Three-Dimensional Bi-Chiral Photonic Crystals," Advanced Materials **21**, 4680-4682 (2009).
24. N. Yu, F. Aieta, P. Genevet, M. A. Kats, Z. Gaburro, and F. Capasso, "A broadband, background-free quarter-wave plate based on plasmonic metasurfaces," Nano Lett. **12**, 6328-6333 (2012).
25. S. Chen, D. Katsis, A. Schmid, J. Mastrangelo, T. Tsutsui, and T. Blanton, "Circularly polarized light generated by photoexcitation of luminophores in glassy liquid-crystal films," Nature **397**, 506-508 (1999).
26. M. Choi, S. H. Lee, Y. Kim, S. B. Kang, J. Shin, M. H. Kwak, K.-Y. Kang, Y.-H. Lee, N. Park, and B. Min, "A terahertz metamaterial with unnaturally high refractive index," Nature **470**, 369-373 (2011).
27. I. V. Lindell, A. Sihvola, S. Tretyakov, and A. Viitanen, *Electromagnetic waves in chiral and bi-isotropic media* (Artech Print on Demand, 1994).
28. S. G. Tikhodeev, A. L. Yablonskii, E. A. Muljarov, N. A. Gippius, and T. Ishihara, "Quasiguided modes and optical properties of photonic crystal slabs," Phys. Rev. B **66** (2002).
29. A. Guo, G. J. Salamo, D. Duchesne, R. Morandotti, M. Volatier-Ravat, V. Aimez, G. A. Siviloglou, and D. N. Christodoulides, "Observation of PT-Symmetry Breaking in Complex Optical Potentials," Phys. Rev. Lett. **103** (2009).
30. J. Shin, J.-T. Shen, and S. Fan, "Three-dimensional metamaterials with an ultrahigh effective refractive index over a broad bandwidth," Phys. Rev. Lett. **102**, 093903 (2009).
31. A. Poddubny, I. Iorsh, P. Belov, and Y. Kivshar, "Hyperbolic metamaterials," Nature Photon. **7**, 948-957 (2013).
32. A. Pors, M. G. Nielsen, G. Della Valle, M. Willatzen, O. Albrektsen, and S. I. Bozhevolnyi, "Plasmonic metamaterial wave retarders in reflection by orthogonally oriented detuned electrical dipoles," Opt. Lett. **36**, 1626-1628 (2011).
33. S.-C. Jiang, X. Xiong, P. Sarriugarte, S.-W. Jiang, X.-B. Yin, Y. Wang, R.-W. Peng, D. Wu, R. Hillenbrand, and X. Zhang, "Tuning the polarization state of light via time retardation with a microstructured surface," Phys. Rev. B **88**, 161104 (2013).
34. X. Piao, S. Yu, J. Hong, and N. Park, "Spectral separation of optical spin based on antisymmetric Fano resonances," Scientific reports **5** (2015).

35. S. Yu, X. Piao, J. Hong, and N. Park, "Bloch-like waves in random-walk potentials based on supersymmetry," Nature Comm. **6** (2015).
36. M.-A. Miri, M. Heinrich, and D. N. Christodoulides, "Supersymmetry-generated complex optical potentials with real spectra," Phys. Rev. A **87**, 043819 (2013).
37. S. Yu, X. Piao, J. Hong, and N. Park, "Metadisorder for designer light in random-walk systems," arXiv preprint arXiv:1510.05518 (2015).

# Low-dimensional optical chirality in complex potentials: supplementary material

Sunkyu Yu,[1,†] Hyun Sung Park,[2,†] Xianji Piao,[1] Bumki Min,[2] Namkyoo Park,[1,*]

[1]*Photonic Systems Laboratory, Department of Electrical and Computer Engineering, Seoul National University, Seoul 08826, Korea*
[2]*Department of Mechanical Engineering, Korea Advanced Institute of Science and Technology (KAIST), Daejeon 305-751, Korea*
*\*Corresponding author: nkpark@snu.ac.kr*
*†These authors contributed equally to this work.*

This document provides supplementary information to "Low-dimensional optical chirality in complex potentials,"

## Supplementary Note 1. Planewave solution of a PT-symmetric optical material

To clarify a point of importance in the chirality of a PT-symmetric optical material, here we assume the simplest case of a planewave propagating along the x-axis ($E = E_0 \cdot e^{i(\omega t - \beta x)}$). From the PT-symmetric permittivity tensor $\varepsilon_r$ of the material, Maxwell's wave equation reduces to the following 2-dimensional matrix equation:

$$\begin{bmatrix} \varepsilon_{\text{eig}} - \varepsilon_{r0} - i\varepsilon_{i0} & -\varepsilon_{\kappa 0} \\ -\varepsilon_{\kappa 0}^* & \varepsilon_{\text{eig}} - \varepsilon_{r0} + i\varepsilon_{i0} \end{bmatrix} \begin{bmatrix} E_y \\ E_z \end{bmatrix} = \mathbf{O}, \quad (1)$$

where $\varepsilon_{\text{eig}}$ is the effective permittivity of the eigenmodes ($\beta^2 = \varepsilon_{\text{eig}} \cdot (\omega/c)^2$). The effective permittivity $\varepsilon_{\text{eig}1,2}$ and the corresponding eigenmode $v_{\text{eig}1,2}$ are then obtained as

$$\varepsilon_{\text{eig}1,2} = \varepsilon_{r0} \pm \sqrt{|\varepsilon_{\kappa 0}|^2 - \varepsilon_{i0}^2}, \quad (2)$$

$$\mathbf{v}_{\text{eig}1,2} = \eta_{1,2} \cdot \begin{bmatrix} \varepsilon_{\kappa 0} \\ -i\varepsilon_{i0} \pm \sqrt{|\varepsilon_{\kappa 0}|^2 - \varepsilon_{i0}^2} \end{bmatrix}, \quad (3)$$

where $\eta_{1,2} = [1/(|\varepsilon_{\kappa 0}|^2 + |-i\varepsilon_{i0} \pm (\varepsilon_{\kappa 0}^2 - \varepsilon_{i0}^2)^{1/2}|^2)]^{1/2}$ is the normalization factor for each eigenmode. To define the exceptional point (EP) originating from the onset of PT symmetry breaking [1,2], we now introduce the interaction parameter [2] $\lambda_{\text{PT}} = (|\varepsilon_{\kappa 0}|^2 - \varepsilon_{i0}^2)^{1/2}$, where $\lambda_{\text{PT}} = 0$ at the EP provides the geometric multiplicity of 1 (= the dimensionality of the eigenspace) for $\varepsilon_{\text{eig}1} = \varepsilon_{\text{eig}2}$. In this case, Supplementary Eqs (2) and (3) are simplified as

$$\varepsilon_{\text{eig}1,2} = \varepsilon_{r0} \pm \lambda_{\text{PT}}, \quad (4)$$

$$\mathbf{v}_{\text{eig}1,2} = \eta_{1,2} \cdot \begin{bmatrix} \varepsilon_{\kappa 0} \\ -i\varepsilon_{i0} \pm \lambda_{\text{PT}} \end{bmatrix}. \quad (5)$$

For reference, from the condition of zero magneto-electric coupling in the present analysis, the value of $\varepsilon_{\kappa 0}$ is real [3] ($\lambda_{\text{PT}} = (\varepsilon_{\kappa 0}^2 - \varepsilon_{i0}^2)^{1/2}$).

## Supplementary Note 2. Effect of imperfect PT symmetry on the modal chirality

For completeness, here we investigate the effect of imperfect PT symmetry on the modal chirality by considering two different situations related to the condition of PT symmetry: (1) non-symmetric real parts ($Re[\varepsilon_y] \neq Re[\varepsilon_z]$) and (2) non-anti-symmetric imaginary parts of permittivity ($Im[\varepsilon_y] \neq -Im[\varepsilon_z]$).

### A. Broken symmetry in the real part of permittivity

To introduce the real-part imperfection ($Re[\varepsilon_y] \neq Re[\varepsilon_z]$), we change Supplementary Eq. (1), including the real part difference $\Delta\varepsilon_{r0}$, as

$$\begin{bmatrix} \varepsilon_{\text{eig}} - \varepsilon_{r0} - \Delta\varepsilon_{r0}/2 - i\varepsilon_{i0} & -\varepsilon_{\kappa 0} \\ -\varepsilon_{\kappa 0}^* & \varepsilon_{\text{eig}} - \varepsilon_{r0} + \Delta\varepsilon_{r0}/2 + i\varepsilon_{i0} \end{bmatrix} \begin{bmatrix} E_y \\ E_z \end{bmatrix} = \mathbf{O}, \quad (6)$$

from $Re[\varepsilon_y] = \varepsilon_{r0} + \Delta\varepsilon_{r0}/2$ and $Re[\varepsilon_z] = \varepsilon_{r0} - \Delta\varepsilon_{r0}/2$. Then, the effective permittivity $\varepsilon_{\text{eig}1,2}$ and the corresponding eigenmode $v_{\text{eig}1,2}$, including the effect of the real part imperfection, are obtained as

$$\varepsilon_{\text{eig}1,2} = \varepsilon_{r0} \pm \sqrt{|\varepsilon_{\kappa 0}|^2 - \varepsilon_{i0}^2 + \frac{\Delta\varepsilon_{r0}}{4}(\Delta\varepsilon_{r0} + 4i\varepsilon_{i0})}, \quad (7)$$

$$\mathbf{v}_{\text{eig}1,2} = \eta_{1,2} \cdot \begin{bmatrix} \varepsilon_{\kappa 0} \\ -\Delta\varepsilon_{r0}/2 - i\varepsilon_{i0} \pm \sqrt{|\varepsilon_{\kappa 0}|^2 - \varepsilon_{i0}^2 + \frac{\Delta\varepsilon_{r0}}{4}(\Delta\varepsilon_{r0} + 4i\varepsilon_{i0})} \end{bmatrix}, \quad (8)$$

where $\eta_{1,2}$ is the new normalization factor for each eigenmode satisfying $|v_{\text{eig}1,2}|^2 = 1$. Due to the complex form inside the square root of Supplementary Eq. (7), perfect coalescence does not occur if $\Delta\varepsilon_{r0} \neq 0$, as shown in Supplementary Figs 1a and 1b. From Supplementary Eq. (8), we can also calculate the modal chirality $\chi_{1,2}$ (see Appendix A), which shows the effect of the imperfection on the modal chirality (Supplementary Fig. 1c). From those results, we can determine the boundary of the tolerance for the modal chirality (Supplementary Fig. 1d).

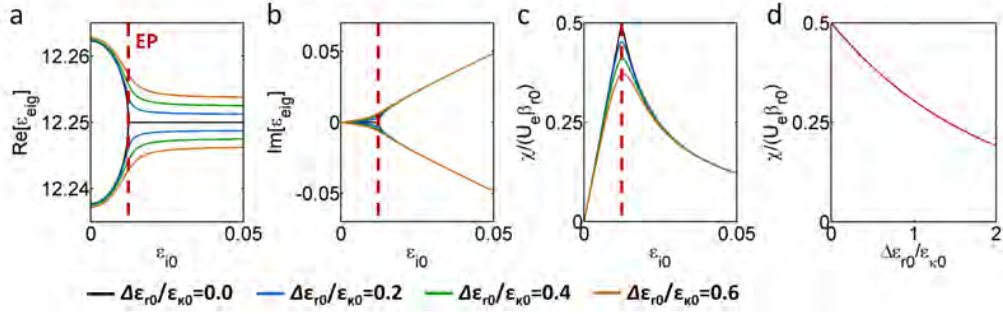

**Supplementary Figure 1. The effect of the imperfection in the real part of PT symmetry** a. The real and b. imaginary parts of the effective permittivity $\varepsilon_{\text{eig}1,2}$. c. The density of chirality $\chi = [\varepsilon_0 \varepsilon_r \cdot E \cdot (\nabla \times E) + (1/\mu_0) \cdot B \cdot (\nabla \times B)]/2$, normalized by the product of the electric field intensity $U_e = |E|^2$ and $\beta_{r0} = \varepsilon_{r0}^{1/2} \cdot 2\pi/\Lambda_0$. d. The density of chirality as a function of the defect $\Delta\varepsilon_{r0}$ at the EP (red dotted lines in a-c). All other parameters are the same as those in Fig. 1 in the main manuscript.

### B. Broken anti-symmetry in the imaginary part of the permittivity

For the imaginary-part imperfection ($Im[\varepsilon_y] \neq -Im[\varepsilon_z]$), Supplementary Eq. (1) is changed, including the new imaginary part $\Delta\varepsilon_{i0}$, as

$$\begin{bmatrix} \varepsilon_{\text{eig}} - \varepsilon_{r0} - i(\varepsilon_{i0} + \Delta\varepsilon_{i0}/2) & -\varepsilon_{\kappa 0} \\ -\varepsilon_{\kappa 0}^* & \varepsilon_{\text{eig}} - \varepsilon_{r0} + i(\varepsilon_{i0} - \Delta\varepsilon_{i0}/2) \end{bmatrix} \begin{bmatrix} E_y \\ E_z \end{bmatrix} = \mathbf{O}, \quad (9)$$

from $Im[\varepsilon_y] = \varepsilon_{i0} + \Delta\varepsilon_{i0}/2$ and $Im[\varepsilon_z] = -\varepsilon_{i0} + \Delta\varepsilon_{i0}/2$, showing the broken anti-symmetry. Then, the effective permittivity $\varepsilon_{\text{eig}1,2}$, including the effect of the imaginary part imperfection, is obtained as

$$\varepsilon_{\text{eig}1,2} = \varepsilon_{r0} + i\frac{\Delta\varepsilon_{r0}}{2} \pm \sqrt{|\varepsilon_{\kappa 0}|^2 - \varepsilon_{i0}^2} = \varepsilon_{r0} + i\frac{\Delta\varepsilon_{r0}}{2} \pm \lambda_{\text{PT}}, \quad (10)$$

while $v_{\text{eig}1,2}$ and the following $\chi_{1,2}$ are the same as Supplementary Eq. (3). The modal chirality is thus not influenced by the imperfect anti-symmetry of the imaginary part. Such a result can be understood within the context of gauge transformation, which has been demonstrated in passive PT-symmetric systems [2,4] ($\Delta\varepsilon_{i0} < 0$).

### Supplementary Note 3. Transfer between RCP and LCP modes in the PT-symmetric chiral material

For an incident wave impinging upon the PT-symmetric material ($E_{\text{inc}} = A_{\text{eig}1} \cdot v_{\text{eig}1} + A_{\text{eig}2} \cdot v_{\text{eig}2}$), the transmitted wave at $x = d$ is expressed as $E_{\text{trn}} = A_{\text{eig}1} \cdot v_{\text{eig}1} \cdot exp(-i\beta_1 \cdot d) + A_{\text{eig}2} \cdot v_{\text{eig}2} \cdot exp(-i\beta_2 \cdot d)$, where $v_{\text{eig}1}$ and $v_{\text{eig}2}$ are complex eigenmodes in Supplementary Eq. (5), $\beta_{1,2} = 2\pi\varepsilon_{\text{eig}1,2}^{1/2}/\Lambda_0$ is the propagation constant of each eigenmode, and $\Lambda_0$ is the free-space wavelength of the wave. To investigate the chiral behavior of the wave, we employ the convenient basis of RCP and LCP: $v_{\text{R,L}} = (1/2)^{1/2} \cdot (1, \pm i)^T$. From the relation of $v_{\text{eig}1,2} = A_{R1,2} \cdot v_R + A_{L1,2} \cdot v_L \equiv [A_{R1,2}, A_{L1,2}]_{CP}^T$, a transfer equation between the incident wave $E_{\text{inc}} = [E_{RI}, E_{LI}]_{CP}^T$ and the transmitted wave $E_{\text{trn}} = [E_{RT}, E_{LT}]_{CP}^T$ in CP bases is obtained as

$$\begin{bmatrix} E_{RT} \\ E_{LT} \end{bmatrix} = \frac{1}{A_{R1}A_{L2} - A_{R2}A_{L1}} \begin{bmatrix} A_{R1}e^{-i\beta_1 d} & A_{R2}e^{-i\beta_2 d} \\ A_{L1}e^{-i\beta_1 d} & A_{L2}e^{-i\beta_2 d} \end{bmatrix} \begin{bmatrix} A_{L2} & -A_{R2} \\ -A_{L1} & A_{R1} \end{bmatrix} \begin{bmatrix} E_{RI} \\ E_{LI} \end{bmatrix}, \quad (11)$$

where the explicit expression for $[A_{R1,2}, A_{L1,2}]_{CP}^T$,

$$\begin{bmatrix} A_{R1,2} \\ A_{L1,2} \end{bmatrix} = \frac{\eta_{1,2}}{\sqrt{2}} \begin{bmatrix} \varepsilon_{\kappa 0} - \varepsilon_{i0} \mp i\lambda_{\text{PT}} \\ \varepsilon_{\kappa 0} + \varepsilon_{i0} \pm i\lambda_{\text{PT}} \end{bmatrix}, \quad (12)$$

is obtained from Supplementary Eq. (5). By applying Supplementary Eq. (12), Supplementary Eq. (11) can be re-expressed using only structural ($d$) and material ($\varepsilon_{r0}$, $\varepsilon_{i0}$, and $\varepsilon_{\kappa 0}$) parameters as $E_{\text{trn}} = M_{\text{PT}} E_{\text{inc}}$ where

$$\mathbf{M}_{\text{PT}} = \begin{bmatrix} t_{R \to R} & t_{L \to R} \\ t_{R \to L} & t_{L \to L} \end{bmatrix}$$
$$= \frac{1}{2}\begin{bmatrix} \varphi_1 + \varphi_2 & -i\frac{\varepsilon_{\kappa 0} - \varepsilon_{i0}}{\lambda_{\text{PT}}}(\varphi_1 - \varphi_2) \\ i\frac{\varepsilon_{\kappa 0} + \varepsilon_{i0}}{\lambda_{\text{PT}}}(\varphi_1 - \varphi_2) & \varphi_1 + \varphi_2 \end{bmatrix}, \quad (13)$$

$\varphi_{1,2} = exp(-i\beta_{1,2}d)$, $d$ is the propagation distance, $\beta_{1,2} = 2\pi\varepsilon_{\text{eig}1,2}^{1/2}/\Lambda_0$, and $\Lambda_0$ is the free-space wavelength. Equation (13) shows the apparent chiral transfer through two unequal off-diagonal terms. The magnitude of the chirality in the intermodal transfer between CP modes is quantified by $C_{\text{IM}}$ as Eq. (2) in the main manuscript.

### Supplementary Note 4. Strength of chiral conversion $C_{CS}$

After the EP ($\varepsilon_{i0} \geq \varepsilon_{\kappa 0}$), $C_{CS}^{R \to L}$ and $C_{CS}^{L \to R}$ can be expressed as

$$C_{CS}^{R \to L} = \left(\frac{\varepsilon_{i0} + \varepsilon_{\kappa 0}}{\varepsilon_{i0} - \varepsilon_{\kappa 0}}\right)^{1/2} \cdot \tanh\left[\frac{2\pi d}{\Lambda_0} \cdot \left(\frac{\sqrt{\varepsilon_{r0}^2 + \varepsilon_{i0}^2 - \varepsilon_{\kappa 0}^2} - \varepsilon_{r0}}{2}\right)^{1/2}\right], \quad (14)$$

$$C_{CS}^{L \to R} = \left(\frac{\varepsilon_{i0} - \varepsilon_{\kappa 0}}{\varepsilon_{i0} + \varepsilon_{\kappa 0}}\right)^{1/2} \cdot \tanh\left[\frac{2\pi d}{\Lambda_0} \cdot \left(\frac{\sqrt{\varepsilon_{r0}^2 + \varepsilon_{i0}^2 - \varepsilon_{\kappa 0}^2} - \varepsilon_{r0}}{2}\right)^{1/2}\right]. \quad (15)$$

When $\varepsilon_{i0} \sim \varepsilon_{\kappa 0}$, $C_{CS}^{R \to L} \sim 2\pi L_{\text{eff}} \cdot (\varepsilon_{i0}/\varepsilon_{r0})$ and $C_{CS}^{L \to R} \sim 0$ near the EP, showing unidirectional chiral conversion (Fig. 1j in the main manuscript).

Because $Re[\beta_1] \neq Re[\beta_2]$ before the EP (Fig. 1a, in the main manuscript), $C_{CS}$ becomes oscillatory, in contrast to the case after the EP ($Re[\beta_1] = Re[\beta_2]$). The $C_{CS}$ ($\varepsilon_{i0} < \varepsilon_{\kappa 0}$) is then expressed as

$$C_{CS}^{R \to L} = \left(\frac{\varepsilon_{\kappa 0} + \varepsilon_{i0}}{\varepsilon_{\kappa 0} - \varepsilon_{i0}}\right)^{1/2} \cdot \left|\tan\left(\frac{2\pi d}{\Lambda_0} \cdot \frac{\sqrt{\varepsilon_{\kappa 0}^2 - \varepsilon_{i0}^2}}{\left(\varepsilon_{r0} + \sqrt{\varepsilon_{\kappa 0}^2 - \varepsilon_{i0}^2}\right)^{1/2} + \left(\varepsilon_{r0} - \sqrt{\varepsilon_{\kappa 0}^2 - \varepsilon_{i0}^2}\right)^{1/2}}\right)\right|, \quad (16)$$

$$C_{CS}^{L \to R} = \left(\frac{\varepsilon_{\kappa 0} - \varepsilon_{i0}}{\varepsilon_{\kappa 0} + \varepsilon_{i0}}\right)^{1/2} \cdot \left| \tan\left( \frac{2\pi d}{\Lambda_0} \cdot \frac{\sqrt{\varepsilon_{\kappa 0}^2 - \varepsilon_{i0}^2}}{\left(\varepsilon_{r0} + \sqrt{\varepsilon_{\kappa 0}^2 - \varepsilon_{i0}^2}\right)^{1/2} + \left(\varepsilon_{r0} - \sqrt{\varepsilon_{\kappa 0}^2 - \varepsilon_{i0}^2}\right)^{1/2}} \right) \right|, \quad (17)$$

which also gives the values of $C_{CS}^{R \to L} = 2\pi L_{eff}(\varepsilon_{i0}/\varepsilon_{r0})$ and $C_{CS}^{L \to R} = 0$ at the EP, equivalent to those of Eqs. (14) and (15).

Supplementary Figure 2 shows the behaviors of $C_{CS}^{R \to L}$ and $C_{CS}^{L \to R}$ with a different $\varepsilon_{\kappa 0}$ as functions of ($\varepsilon_{i0}/\varepsilon_{\kappa 0}$) before the EP ($\varepsilon_{i0} < \varepsilon_{\kappa 0}$) and after the EP ($\varepsilon_{i0} > \varepsilon_{\kappa 0}$). From the unequal strength of $C_{CS}^{R \to L} > C_{CS}^{L \to R}$, there exists an LCP-favored chiral conversion before the EP, although it is much weaker than that at the EP. It is noted that the oscillatory behavior of $C_{CS}$ before the EP is determined by $\varepsilon_{\kappa 0}$.

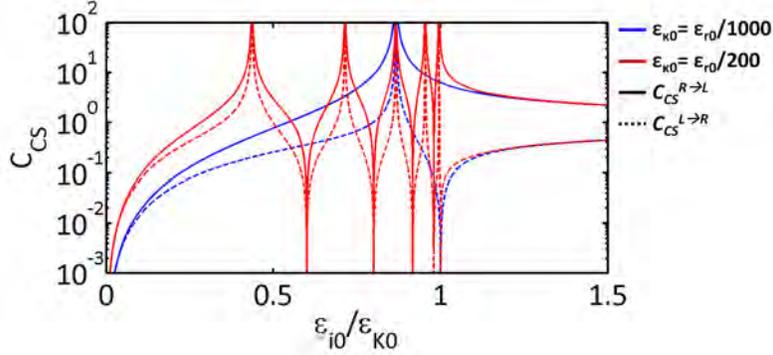

**Supplementary Figure 2.** $C_{CS}^{R \to L}$ **and** $C_{CS}^{L \to R}$ **as functions of ($\varepsilon_{i0}/\varepsilon_{\kappa 0}$)** for the cases of $\varepsilon_{\kappa 0} = \varepsilon_{r0}/1000$ and $\varepsilon_{\kappa 0} = \varepsilon_{r0}/200$. $\varepsilon_{r0} = 6.5$ and $L_{eff} = 10^3$.

## Supplementary Note 5. The state of polarization (SOP) at the EP: Optical spin black hole

For the graphical representation of the SOP, here we apply the Stokes parameter expression (of $S_0$, $S_1$, $S_2$, and $S_3$) on the Poincaré sphere, which follows the relations below:

$$\begin{aligned} S_0 &= |E_x|^2 + |E_y|^2, \\ S_1 &= |E_x|^2 - |E_y|^2, \\ S_2 &= 2\,\text{Re}[E_x^* E_y], \\ S_3 &= 2\,\text{Im}[E_x^* E_y], \end{aligned} \quad (18)$$

where $S_0$ is the radius of the Poincaré sphere, and ($S_1$, $S_2$, $S_3$) is the Cartesian coordinate of the SOP on the sphere. To describe the general tendency of the SOP at the EP, we assume the 400 randomly polarized incidences ($E_x = a + bi$ and $E_y = c + di$ for the incidence where $a$, $b$, $c$, and $d$ are real numbers with the uniform random distribution in [-1,1]) on the PT-symmetric optical potential for Fig. 2c in the main manuscript.

## Supplementary Note 6. Low-dimensional linear polarization

Because the formulation of Supplementary Eq. (1) is based on the general framework for two-level PT-symmetric potentials [2,4,5], our analysis of the low-dimensional polarization can be extended beyond the case of chirality treated in the main manuscript. For example, instead of mixing amplifying $y$-LP and decaying $z$-LP modes, consider the mixing of amplifying RCP and decaying LCP modes, which is possible with the recent development of active chiral materials [6,7] and circular dichroism [8,9]. The eigenmodes of the PT-symmetric material are then expressed as $v_{eig1,2} = \eta_{1,2} \cdot \{\varepsilon_{\kappa 0}\cdot[1,i] + (-i\varepsilon_{i0} \pm \lambda_{PT})\cdot[1,-i]\}^T = \eta_{1,2} \cdot [\varepsilon_{\kappa 0} - i\varepsilon_{i0} \pm \lambda_{PT}, -i\cdot(-\varepsilon_{\kappa 0} - i\varepsilon_{i0} \pm \lambda_{PT})]^T$, while the eigenvalues are the same as in Figs 1a and 1b in the main manuscript.

Supplementary Figure 3 shows the corresponding profiles of the eigenpolarizations for points $d$-$h$. For the Hermitian case (point $d$), the eigenmodes are linearly polarized (LP) due to the even and odd couplings of the RCP and LCP modes. When $\varepsilon_{i0}$ increases (0 < $\varepsilon_{i0} < \varepsilon_{\kappa 0}$, point $e$), the eigenmodes begin to converge. At the EP ($\varepsilon_{i0} = \varepsilon_{\kappa 0}$, point $f$), two LP eigenmodes have coalesced, and the reduction to a *1-dimensional LP basis* is evident. After the EP ($\varepsilon_{i0} > \varepsilon_{\kappa 0}$, points $g$ and $h$), each eigenmode is saturated to a CP mode (RCP, amplifying, and LCP, decaying).

Because a low-dimensional eigenstate can be designed deliberately in the form both of CP (main manuscript, poles on the Poincaré sphere) and LP modes (Supplementary Fig. 3, the equator on the Poincaré sphere), all of the states on the Poincaré sphere can be designed as the low-dimensional EP, by combining the linear loss and circular dichroism. Therefore, the dynamics of optical spin black hole can be achieved on the entire Poincaré sphere.

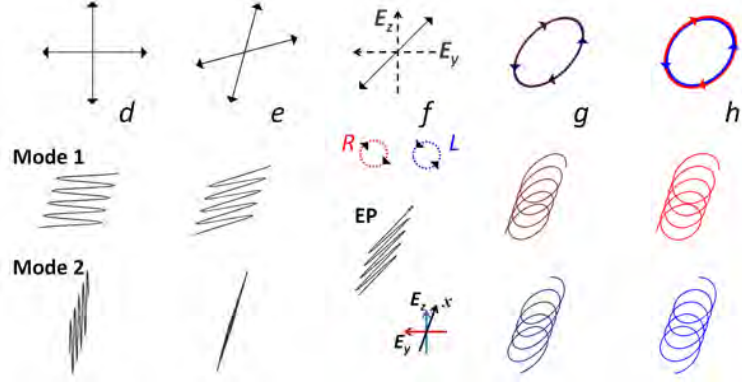

**Supplementary Figure 3. Spatial evolutions of eigenmodes in the low-dimensional LP material corresponding to points** *d-h* At the exceptional point *f*, the complex eigenmode has the singular form of an LP state (*y–z*). $\varepsilon_{r0}$ = 12.25, and $\varepsilon_{k0} = \varepsilon_{r0}/10^3 > 0$.

## Supplementary Note 7. Chirality reversal for LP incidences

Focusing on the vicinity of the EP, chirality reversal can be achieved for the SOP of linear polarization (LP) incidences. For the LP incidence $E_{inc} = (E_{RI}, E_{LI})_{CP}^T = (1/2)^{1/2} \cdot E_0 \cdot (e^{-i\theta}, e^{i\theta})_{CP}^T$ of an arbitrary polarized angle $\theta$, the transmitted field $E_{trn}$ at the EP is obtained from the transfer matrix $M_{PT}$ as

$$\mathbf{E}_{trn} = \begin{pmatrix} E_{RT} \\ E_{LT} \end{pmatrix} = \frac{E_0}{\sqrt{2}} \cdot e^{-i \cdot 2\pi \frac{\sqrt{\varepsilon_{r0}} \cdot d}{\Lambda_0}} \cdot \begin{pmatrix} e^{-i\theta} \\ e^{i\theta} + C_{CS}^{R \to L}|_{EP} \cdot e^{-i\theta} \end{pmatrix}. \quad (19)$$

From Supplementary Eq. (19), it is clear that LCP transmission can be controlled by the polarized angle $\theta$ of LP and $C_{CS}^{R \to L}|_{EP}$ (Supplementary Fig. 4a), whereas RCP transmission is invariant (see Supplementary Movie 2 for the SOP evolutions by the degree of PT symmetry with LP incidences; $\theta$ = 0, $\pi/2$, $\pi$, and $3\pi/2$). In the strong chiral conversion regime of $C_{CS}^{R \to L}|_{EP} > 2$ (outside the dotted circle in Supplementary Fig. 4a), the transmitted wave is *always* left-handed ($|E_{LT}| > |E_{RT}|$) for all angles of $\theta$. In contrast, in the regime of $C_{CS}^{R \to L}|_{EP} \leq 2$, *chirality reversal* to a right-handed output is permitted for input angles of $cos(2\theta) < -C_{CS}^{R \to L}|_{EP}/2$. A pure RCP transmission can also be achieved in the special case of $C_{CS}^{R \to L}|_{EP} = 1$ and z-LP ($\theta$ = 90°, on the solid circle in Supplementary Fig. 4a), which is counterintuitive regarding the singular existence of the LCP modal helix. This paradoxical result arises from the unidirectional ($R \to L$) intermodal transfer, which leads to a completely destructive interference for the LCP mode only ($e^{i\theta} + C_{CS}^{R \to L}|_{EP} \cdot e^{-i\theta}$). Note that the observed phenomena of CP interference and chirality reversal are absent in conventional chiral materials [10] which are based on *uncoupled* LCP and RCP modes ($t_{L \to R} = t_{R \to L} = 0$), and thus, they maintain the LP state with natural optical rotation during propagation, *a.k.a.* optical activity. In practice, the regime of chirality reversal can be controlled by changing $L_{eff}$ or $\varepsilon_{k0}$ (Supplementary Fig. 4b) based on the definition of $C_{CS}^{R \to L}|_{EP}$ (=$2\pi L_{eff} \cdot (\varepsilon_{k0}/\varepsilon_{r0})$).

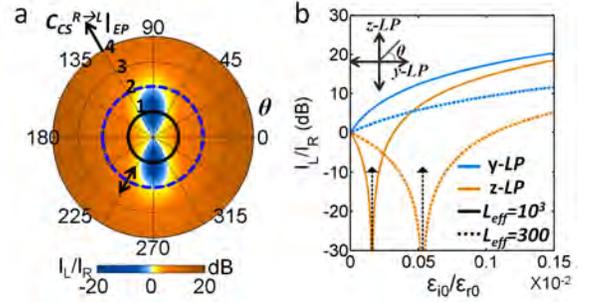

**Supplementary Figure 4. Chirality reversal for LP incidences** a. The ratio of $I_L/I_R$ at the EP for LP incidences is shown in a as a function of $C_{CS}^{R \to L}|_{EP}$ and the polarized angle $\theta$ of LP and in b as a function of coupling permittivity ($\varepsilon_{k0}/\varepsilon_{r0}$). The black dotted arrows denote the points at which $C_{CS}^{R \to L}|_{EP} = 1$. All the results are based on the transfer matrix method. $\varepsilon_{r0}$=6.5 and $L_{eff}$=$10^3$ for a.

## Supplementary Note 8. Realization of PT-symmetric permittivity in metamaterial platforms

To transplant the point-wise anisotropic permittivity of a PT-symmetric chiral material into the structure composed of isotropic materials, we first consider a metamaterial platform in a THz regime. As a unit element, we adopt the I-shaped patch (Supplementary Fig. 5a) for each polarization, which induces the enhanced light-matter interaction through the deep subwavelength (~$\lambda_0$/200) gap structure [11]. By crossing the *y*- and *z*-unit elements (Supplementary Fig. 5b), we obtain the metamaterial, which supports the well-known electrical response of the Lorentz model both for *y*- and *z*-polarizations, as

$$\varepsilon_{ry}(\omega) = \varepsilon_{poly} + \frac{\omega_{py}^2}{\omega_{0y}^2 - \omega^2 + i\gamma_y \omega}, \quad \varepsilon_{rz}(\omega) = \varepsilon_{poly} + \frac{\omega_{pz}^2}{\omega_{0z}^2 - \omega^2 + i\gamma_z \omega} \quad (20)$$

where $\varepsilon_{poly}$ is the permittivity of the polyimide ($\varepsilon_{poly}$ = 3.238 – 0.144*i*), $\omega_{0(y,z)}$ is the characteristic frequency, $\gamma_{(y,z)}$ is the damping coefficient, and $\omega_{p(y,z)}$ is the plasma frequency of the metamaterial for *y*- and *z*-polarizations. Because the near-field intensity distribution is widely separated for each polarization (Supplementary Figs 5c and 5d), it is worth mentioning that the Lorentz permittivity curve of each polarization can be detuned independently to realize the required anisotropic metamaterial.

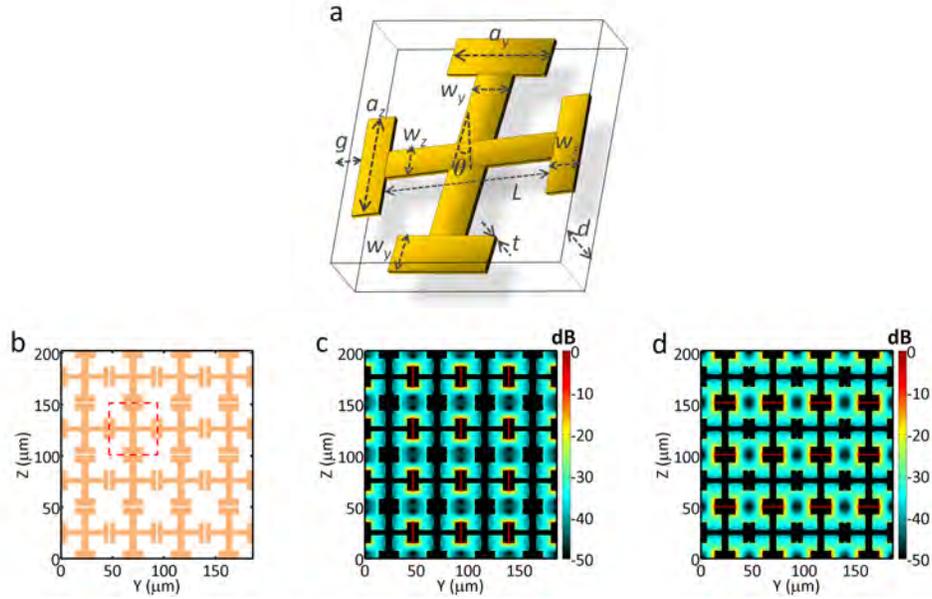

**Supplementary Figure 5. Crossed I-shaped metamaterials** a. Metamaterial realization of Fig. 4a in the main manuscript using crossed I-shaped gold structures (tilted by $\theta$) inside the polyimide. b. 2D cross-sectional schematic view of the structure composed of gold (yellow) inside the polyimide (white). The tilted angle between patches is $\theta = 0°$, and all the other parameters are the same as those in Fig. 4 in the main manuscript. The red dashed box represents the unit-cell metamaterial. The electrical intensity distribution calculated by COMSOL ($|E|^2$ in dB scale) is shown in c. for a $y$-polarized and d. a $z$-polarized planewave incidence at the peak frequency of the permittivity curves in Fig. 4c in the main manuscript.

As noted in the main manuscript, we utilize the low-loss ($|Re[\varepsilon]| \gg |Im[\varepsilon]|$) and high-loss ($|Re[\varepsilon]| \sim |Im[\varepsilon]|$) dielectric regimes of the Lorentz response to obtain the passive PT symmetry condition of $Re[\varepsilon_y] = Re[\varepsilon_z]$ and $Im[\varepsilon_z] < Im[\varepsilon_y] < 0$ (see Supplementary Note 2). Because the modal chirality is critically dependent on the condition of $Re[\varepsilon_y] = Re[\varepsilon_z]$ (see Supplementary Note 2), we need to intervein the real part of the Lorentz curves. To achieve this, the effect of varying the physical quantities of the Lorentz model ($\omega_0$, $\omega_p$, and $\gamma$) is shown in Supplementary Fig. 6. Also corresponding to the Supplementary Eq. (20), it is necessary to change the characteristic frequency of $\omega_0$ for the spectral shift (Supplementary Fig. 6a), while the magnitude of the permittivity is mainly tuned with the plasma frequency $\omega_p$ (Supplementary Fig. 6b compared to 6c).

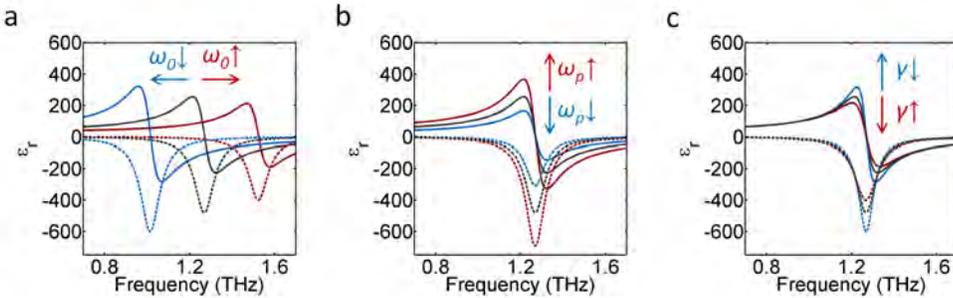

**Supplementary Figure 6. The effect of changing the physical quantities ($\omega_0$, $\omega_p$, and $\gamma$) in the Lorentz model** The permittivities $\varepsilon(\omega)$ are shown for the changes of a. the characteristic frequency $\omega_0$ (black: $\omega_{00}$, red: $1.2\cdot\omega_{00}$, blue: $0.8\cdot\omega_{00}$), b. the plasma frequency $\omega_p$ (black: $\omega_{p0}$, red: $1.2\cdot\omega_{p0}$, blue: $0.8\cdot\omega_{p0}$), and c. the damping coefficient $\gamma$ (black: $\gamma_0$, red: $1.2\cdot\gamma_0$, blue: $0.8\cdot\gamma_0$). $\omega_{00} = 1.27$THz, $\omega_{p0} = 8.2$THz, and $\gamma_0 = 0.11$THz. Solid (dotted) lines denote the $Re[\varepsilon]$ ($Im[\varepsilon]$).

By considering the results of Supplementary Fig. 6, now we investigate the role of the structural parameters ($a$, $g$, $L$, and $w$ of Supplementary Fig. 5a) for the design of PT-symmetric permittivity relating to the physical quantities ($\omega_0$, $\omega_p$, and $\gamma$). To satisfy the condition of $Re[\varepsilon_y] = Re[\varepsilon_z]$ and $Im[\varepsilon_z] < Im[\varepsilon_y] < 0$ with two Lorentz curves in the dielectric regime, it is necessary to introduce the frequency shift between these curves by designing different characteristic frequencies of $\omega_{01} \neq \omega_{02}$ for I-shaped patches 1 and 2 (1, 2 represent $y$ or $z$). To observe the EP with the spectral stability, we consider only the cases of the broadband design of EP (black circles in Supplementary Figs 7a,7b), neglecting cases that are too sensitive (green circles in Supplementary Figs 7a,7b). If I-shaped patch 1 in the high-frequency regime ($\omega_{01}$) supports a lower plasma frequency than that of I-shaped patch 2 ($\omega_{p1} < \omega_{p2}$, Supplementary Fig. 7a), the condition with broad bandwidth is satisfied at the metallic state (Fig. 4e in the main manuscript), which supports an evanescent mode inside the PT-symmetric metamaterial. Meanwhile, if the design satisfying $\omega_{01} > \omega_{02}$ and $\omega_{p1} > \omega_{p2}$ is applied simultaneously (Supplementary Fig. 7b), the propagating mode can be utilized to observe the low-dimensionality (Fig. 4c in the main manuscript).

As shown in Supplementary Figs 7c-7d, increasing the parameters related to the 'length' of the patch ($L$ and $a$) increases the plasma frequency $\omega_p$ due to the large number of participating electrons (note that $\omega_p^2 \sim Ne^2/(m\varepsilon_0)$, where $N$ is the density of electrons), while the characteristic frequency decreases due to the weak restoring force. Similarly, decreasing gap $g$ (Supplementary Fig. 7e) increases the attractive force, yielding the reversed response of $\omega_0\downarrow$ and $\omega_p\uparrow$. Therefore, the strategy for low-dimensional evanescent waves (Supplementary Fig. 7a) can be achieved with the manipulation of $L$, $a$, or $g$. However, by widening the patches (Supplementary Fig. 7f), we can increase the participating electrons ($\omega_p\uparrow$) and the restoring force ($\omega_0\uparrow$) simultaneously, enabling the case of Supplementary Fig. 7b, with low-dimensional propagating waves. Now, by changing the tilted angle $\theta$ (Supplementary Fig. 5a), we can obtain the low-dimensional chiral dynamics at the EP by obtaining the coupling between the $y$- and $z$-polarizations for both cases.

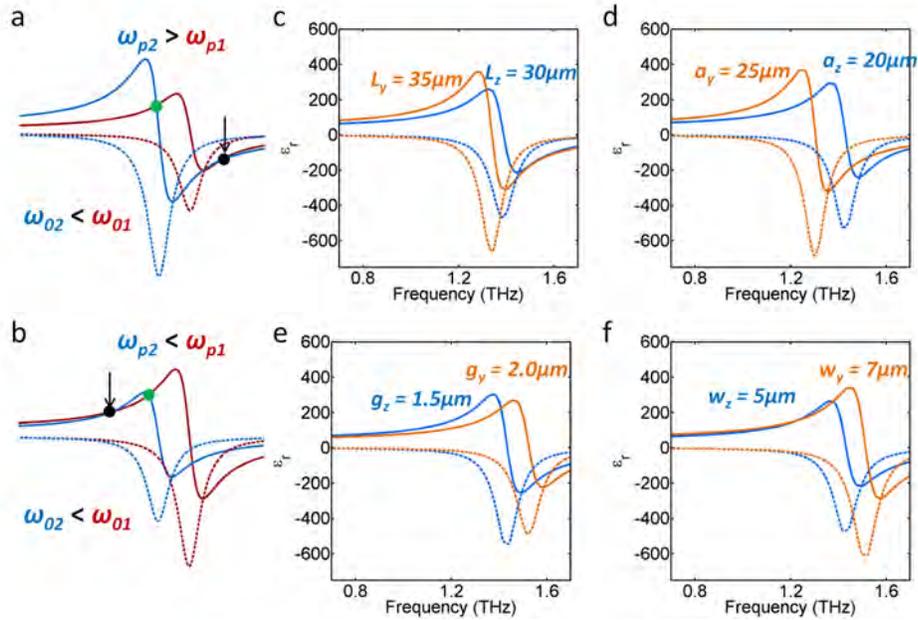

**Supplementary Figure 7. The effect of changing the structural parameters** The strategy for satisfying the condition of PT symmetry with a. $\omega_{01} > \omega_{02}$ and $\omega_{p1} < \omega_{p2}$ (low-dimensional evanescent waves) and b. $\omega_{01} > \omega_{02}$ and $\omega_{p1} > \omega_{p2}$ (low-dimensional propagating waves). Black (or green) circles denote the condition of low-dimensional EP with broad (or narrow) bandwidths. The COMSOL-calculated permittivities $\varepsilon(\omega)$ are shown for the changes of c. length $L$, d. arm length $a$, e. gap thickness $g$, and f. width $w$. The unchanged parameters are $L = 30$ μm, $a = 20$ μm, $g = 1.5$ μm, and $w = 5$ μm.

### References

1. C. M. Bender, and S. Boettcher, "Real spectra in non-Hermitian Hamiltonians having P T symmetry," Phys. Rev. Lett. **80**, 5243 (1998).
2. S. Yu, X. Piao, D. R. Mason, S. In, and N. Park, "Spatiospectral separation of exceptional points in PT-symmetric optical potentials," Phys. Rev. A **86**, 031802 (2012).
3. D. Martin, K. Neal, and T. Dean, "The optical and magneto-optical behaviour of ferromagnetic metals," Proceedings of the Physical Society **86**, 605 (1965).
4. A. Guo, G. J. Salamo, D. Duchesne, R. Morandotti, M. Volatier-Ravat, V. Aimez, G. A. Siviloglou, and D. N. Christodoulides, "Observation of PT-Symmetry Breaking in Complex Optical Potentials," Phys. Rev. Lett. **103** (2009).
5. C. E. Rüter, K. G. Makris, R. El-Ganainy, D. N. Christodoulides, M. Segev, and D. Kip, "Observation of parity–time symmetry in optics," Nature Phys. **6**, 192-195 (2010).
6. S. Furumi, and N. Tamaoki, "Glass-forming cholesteric liquid crystal oligomers for new tunable solid-state laser," Adv Mater **22**, 886-891 (2010).
7. S. Chen, D. Katsis, A. Schmid, J. Mastrangelo, T. Tsutsui, and T. Blanton, "Circularly polarized light generated by photoexcitation of luminophores in glassy liquid-crystal films," Nature **397**, 506-508 (1999).
8. Z. Fan, and A. O. Govorov, "Plasmonic circular dichroism of chiral metal nanoparticle assemblies," Nano Lett **10**, 2580-2587 (2010).
9. C. Song, M. G. Blaber, G. Zhao, P. Zhang, H. C. Fry, G. C. Schatz, and N. L. Rosi, "Tailorable plasmonic circular dichroism properties of helical nanoparticle superstructures," Nano Lett. **13**, 3256-3261 (2013).
10. I. V. Lindell, A. Sihvola, S. Tretyakov, and A. Viitanen, *Electromagnetic waves in chiral and bi-isotropic media* (Artech Print on Demand, 1994).
11. M. Choi, S. H. Lee, Y. Kim, S. B. Kang, J. Shin, M. H. Kwak, K.-Y. Kang, Y.-H. Lee, N. Park, and B. Min, "A terahertz metamaterial with unnaturally high refractive index," Nature **470**, 369-373 (2011).